\documentclass{article}

 \usepackage{tcolorbox}
 \usepackage{booktabs}
 \usepackage{verbatim}
 \usepackage{placeins}
 \usepackage{float}
  \usepackage[preprint]{neurips_2025}

\usepackage[utf8]{inputenc} % allow utf-8 input
\usepackage[T1]{fontenc}    % use 8-bit T1 fonts
\usepackage{hyperref}       % hyperlinks
\usepackage{url}            % simple URL typesetting
\usepackage{booktabs}       % professional-quality tables
\usepackage{amsfonts}       % blackboard math symbols
\usepackage{nicefrac}       % compact symbols for 1/2, etc.
\usepackage{microtype}      % microtypography
\usepackage{xcolor}         % colors
\usepackage{graphicx}
\usepackage{standalone}
\usepackage{booktabs,tabularx,array,makecell}
\usepackage{braket}

\usepackage{markdown}
\usepackage{listings}
\lstdefinestyle{markdownwrap}{
  basicstyle=\ttfamily\fontsize{6}{7}\selectfont,
  breaklines=true,
  breakatwhitespace=true,
  showstringspaces=false,
  columns=fullflexible,
  postbreak=\mbox{},
  breakindent=0pt
}
\usepackage{courier} % For Consolas-like font
\definecolor{codegray}{rgb}{0.95,0.95,0.95}
\definecolor{keywordcolor}{rgb}{0.0,0.0,0.6}
\definecolor{commentcolor}{rgb}{0.0,0.5,0.0}
\definecolor{stringcolor}{rgb}{0.58,0.0,0.82}
\lstdefinestyle{mypython}{
    backgroundcolor=\color{codegray},
    commentstyle=\color{commentcolor}\itshape,
    keywordstyle=\color{keywordcolor}\bfseries,
    stringstyle=\color{stringcolor},
    basicstyle=\fontfamily{pcr}\selectfont\tiny, % Courier = Consolas-like
    breaklines=true,
    frame=single,
    language=Python,
    showstringspaces=false,
    tabsize=4,
    numbers=left,
    numberstyle=\tiny\color{gray},
    numbersep=8pt
}

\usepackage{xspace}
\newcommand{\passatk}{\texttt{pass@k}\xspace}

\definecolor{tickgreen}{HTML}{228B22} % custom green, no dvipsnames needed
\usepackage{pifont}
\newcommand{\tick}{\textcolor{tickgreen}{\ding{51}}}
\newcommand{\blank}{\phantom{\tick}}

\newcommand{\modelname}[1]{{#1}\xspace}
\newcommand{\qwen}{\modelname{Qwen2.5-Coder-14B}}
\newcommand{\qweni}{\modelname{Qwen2.5-Coder-14B-Instruct}}
\newcommand{\qwenqki}{\modelname{Qwen2.5-Coder-14B-Qiskit-Instruct}}
\newcommand{\qwenqkis}{\modelname{Qwen2.5-Coder-14B-Qiskit-Instruct-SFT}}
\newcommand{\qwenqkisg}{\modelname{Qwen2.5-Coder-14B-Qiskit-Instruct-SFT-GRPO}}
\newcommand{\qwenqkid}{\modelname{Qwen2.5-Coder-14B-Qiskit-Instruct-DPO}}
\newcommand{\qwenqkidg}{\modelname{Qwen2.5-Coder-14B-Qiskit-Instruct-DPO-GRPO}}

\newcolumntype{Y}{>{\centering\arraybackslash}X}

\usepackage{ifthen}     % simple conditionals

\definecolor{TodoBG}{HTML}{FFF7D6}   % soft warm background
\definecolor{TodoFG}{HTML}{3A3A3A}   % dark gray text
\definecolor{TodoBD}{HTML}{D8C56A}   % muted border

\definecolor{FixBG}{HTML}{FDE7EA}    % soft red/rose
\definecolor{FixFG}{HTML}{3A3A3A}
\definecolor{FixBD}{HTML}{D88A8A}

\definecolor{NoteBG}{HTML}{E9F2FF}   % soft blue
\definecolor{NoteFG}{HTML}{2E2E2E}
\definecolor{NoteBD}{HTML}{7FA6D9}

\tcbset{colback=blue!5!white, colframe=blue!75!black, 
        boxrule=0.5mm, arc=4mm, outer arc=4mm}
\definecolor{lightblue}{rgb}{0.8,0.9,1}
\definecolor{lightred}{rgb}{1,0.8,0.8}

\newif\ifshowtodos
\showtodostrue        % <— change to \showtodosfalse for "final" mode

\newcommand{\todochip}[4]{%
  \ifshowtodos
    \begingroup
      \setlength{\fboxsep}{1pt}% padding inside the box
      {\sffamily\footnotesize % compact, clean
        \fcolorbox{#3}{#2}{%
          \textcolor{#1}{\textbf{#4}}%
        }%
      }%
    \endgroup
  \fi
}

\newcommand{\todo}[2][]{%
  \ifshowtodos
    \todochip{TodoFG}{TodoBG}{TodoBD}{%
      \ifthenelse{\equal{#1}{}}{}{#1:\,}#2%
    }%
  \fi
}

\newcommand{\todomargin}[2][]{%
  \ifshowtodos
    \marginpar{\raggedright\footnotesize\sffamily
      \setlength{\fboxsep}{1pt}%
      \fcolorbox{TodoBD}{TodoBG}{%
        \parbox{2.5cm}{%
          \textcolor{TodoFG}{%
            \ifthenelse{\equal{#1}{}}{\textbf{TODO}}{\textbf{#1}}:\ %
            #2%
          }%
        }%
      }%
    }%
  \fi
}

\usepackage{mdframed}
\usepackage{amsthm}
\newtheoremstyle{definition-style}%
  {\topsep}{\topsep}%
  {\normalfont}% Body font
  {}% Indent amount
  {\bfseries}{. }% Head font
  {\newline}% Space after theorem head
  {}% Theorem head spec (can also be empty for default)

\theoremstyle{definition-style}
\newtheorem{definition}{Definition}

\newenvironment{definitionbox}
  {\begin{mdframed}\begin{definition}}
  {\end{definition}\end{mdframed}}

\hypersetup{
     colorlinks = true,
     linkcolor = red,
     anchorcolor = blue,
     citecolor = teal,
     filecolor = blue,
     urlcolor = black
     }
\title{Quantum Verifiable Rewards \\ for Post-Training Qiskit Code Assistant}

\author{
 Nicolas Dupuis$^\star$, Adarsh Tiwari$^\star$, Youssef Mroueh$^{\dagger}$,\\ \textbf{David Kremer$^\star$, Ismael Faro$^\star$, Juan Cruz-Benito$^\star$}
  \\
  ~~\\
 $^\star$ IBM Quantum, $^\dagger$ IBM Research \\
}

\begin{document}

\maketitle

\begin{abstract}
Qiskit is an open-source quantum computing framework that allows users to design, simulate, and run quantum circuits on real quantum hardware. We explore post-training techniques for LLMs to assist in writing Qiskit code. We introduce quantum verification as an effective method for ensuring code quality and executability on quantum hardware. To support this, we developed a synthetic data pipeline that generates quantum problem–unit test pairs and used it to create preference data for aligning LLMs with DPO. Additionally, we trained models using GRPO, leveraging quantum-verifiable rewards provided by the quantum hardware. Our best-performing model, combining DPO and GRPO, surpasses the strongest open-source baselines on the challenging Qiskit-HumanEval-hard benchmark.

\end{abstract}

\section{Introduction}
Large language models (LLMs) have demonstrated remarkable capabilities across diverse domains, with their effectiveness largely dependent on post-training procedures that align model behavior with desired outcomes. Traditional approaches such as Reinforcement Learning from Human Feedback (RLHF) \citep{ouyang2022training, christiano2017deep} and more recent innovations like Direct Preference Optimization (DPO) \citep{rafailov2023direct} have primarily relied on human-derived reward signals to guide model refinement. These methods have proven highly effective for both general language tasks and complex reasoning problems. 

Beyond human feedback, in applications such as AI for code, execution-based feedback became the de-facto reward for training such models  within a reinforcement learning loop in order to produce reliable code \citep{dong2024self}. This is  achieved via compiling and executing generated code on CPU and comparing it against a battery of unit tests, that can be used as a feedback to improve the model. This feedback by execution    has been adopted in many closed and open-source models to cite a few 
 DeepSeek-R1 \citep{guo2025deepseek}, Qwen \citep{yang2025qwen3}, Qwen3Coder 
\citep{qwen3coder2025}  and Open-R1 \citep{Openr12025}. Group Relative Policy Optimization (GRPO) \citep{shao2024deepseekmath} allows to integrate efficiently this execution feedback in an online fashion in reinforcement learning. Another approach  would be via creating preference datasets from the execution feedback and use it in an offline fashion within the DPO paradigm.

While these recent advances have focused on verification via execution on a classical computer, we explore in this paper post-training LLMs from  direct feedback from Quantum Processing Units (QPUs) or simulators to generate Qiskit code that is executable on a Quantum Computer. In this context the correctness and efficiency of generated code or circuit designs can only be verified through actual execution on quantum hardware, or realistic quantum simulators.

While hardware-in-the-loop methods have shown promise in robotics and control \citep{liu2022robot,mihalivc2022hardware}, they remain unexplored for optimizing language models. Recent work such as Ether0 \citep{narayanan2025training} demonstrates the potential of reinforcement learning with verifiable rewards to train reasoning models for complex scientific tasks, in this case chemical design, without requiring additional domain pretraining. Ether0’s results highlights how RL-based post-training can leverage structured, verifiable objectives to improve reasoning and data efficiency in domains where correctness can be objectively measured. Quantum computing offers a similar opportunity: unlike traditional language tasks, quantum problems provide objective, measurable outcomes through execution on quantum hardware. Yet, no systematic approach currently incorporates QPU feedback into the post-training optimization loop for LLMs specialized in quantum computing.

This work addresses these limitations by introducing a novel framework for post-training LLMs using reward signals computed directly from quantum processing units. We demonstrate how quantum systems and quantum simulators-related information, configuration and properties can be systematically incorporated into DPO and GRPO optimization procedures, enabling the development of LLMs that are not merely fine-tuned on quantum computing datasets, but are actively optimized based on their performance and reliability when executed on actual quantum systems. Our approach bridges the gap between theoretical quantum computing knowledge and practical implementation capabilities, while contributing new insights into reward-based optimization for physically-grounded AI systems.

Our main contributions are:
\begin{itemize}
\item We introduce quantum verification as a method to ensure code quality and executability on quantum systems (Section \ref{sec:q_verification}).
\item We develop a synthetic data pipeline that generates quantum problem / unit test pairs to support model alignment (Section~\ref{sec:q_verification}).
\item We generate quantum-verified data from synthetic prompts to post-train LLMs using DPO
and applied GRPO using verifiable rewards obtained from quantum hardware (Section~\ref{sec:data-model}).
\item We demonstrate a hybrid model combining DPO and GRPO, which outperforms 
existing open-source baselines on the Qiskit-HumanEval-hard benchmark (Section~\ref{sec:results}).
\end{itemize}

\section{Quantum Verification}
\label{sec:q_verification}
LLMs excel at understanding and writing code in many programming languages. They can ingest in their context a full code base
and be used in complex agentic workflows~\citep{jimenez2024swebench}.
Writing code for quantum is more difficult. One reason is that open-source libraries for quantum computing 
such as Qiskit~\citep{javadiabhari2024quantumcomputingqiskit} evolve fast and deprecate features quite often. This fast pace is in part dictated by the constant progress
made in quantum hardware and quantum computing platforms requiring the Software Development Kits (SDKs) to adapt.
Another challenge is that the programmer, human being or AI, needs to have a good understanding
of the physics behind the quantum job to run, but also being aware of the quantum hardware.
We argue that for those programmers, the best way to write correct quantum computing code
is to directly interact with a quantum system. 
For an AI, which is our main interest in this discussion, the interaction
can happen offline, in which case a model is trained on a dataset with code samples that have been previously 
``verified'' on quantum hardware. The interaction can also happen online by training a policy using reinforcement learning 
to maximize a verifiable reward signal given directly by the quantum hardware. 
In this section we first discuss quantum verification
in the context of Qiskit, then present our data-driven approach for offline and online interaction of an LLM with a quantum system.

\subsection{Qiskit SDK and Qiskit Runtime}

\begin{figure}[t]
\begin{minipage}[h]{0.48\textwidth}
\scalebox{0.8}{
\begin{lstlisting}[style=mypython]
from qiskit import QuantumCircuit
from qiskit_ibm_runtime import Sampler
from qiskit_ibm_runtime import QiskitRuntimeService
from qiskit.transpiler.preset_passmanagers import generate_preset_pass_manager

# 2 qubits, 2 classical bits
bell_circuit = QuantumCircuit(2, 2)
# Hadamard on qubit 0, CNOT control 0, target 1
bell_circuit.h(0)
bell_circuit.cx(0, 1)
# Measure both qubits
bell_circuit.measure([0, 1], [0, 1])
# Initialize Qiskit Runtime Service
service = QiskitRuntimeService()
# Select a suitable backend
backend = service.least_busy(simulator=False)
# Transpile the circuit for the chosen backend
pm = generate_preset_pass_manager(backend=backend)
isa_bell_circuit = pm.run(bell_circuit)
# run the sampler primitive on backend
sampler = Sampler(mode=backend)
job = sampler.run([isa_bell_circuit])
result = job.result()
\end{lstlisting}
}
\end{minipage}%
\hspace{-1cm}
\begin{minipage}[h]{0.5\textwidth}
\includegraphics[width=1.1\textwidth]{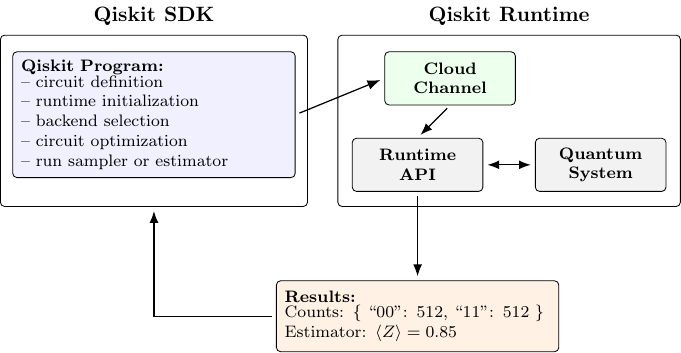}
\end{minipage}
\caption{(--left) Example of a Qiskit program; (--right) Simplified interaction workflow between the Qiskit SDK and the Qiskit Runtime.}
\label{fig:qk-runtime}
\end{figure}

A Qiskit program usually begins by instantiating a \texttt{QuantumCircuit} class, followed by adding gates and measurement operations to it, mapping in this way the actual problem to be solved.
The circuit then needs to be optimized (transpilation) for the coupling map defining
the physical connectivity of the qubits inside the quantum system and its features (noise, configuration, properties). 
For running a circuit,
Qiskit provides the Runtime API accessed through a cloud channel that is responsible for 
submitting the job.
The users have access to two primitives that can be called to run a quantum work through the Qiskit Runtime. 
The estimator primitive computes expectation values $\bra{\Psi}H\ket{\Psi}$ for given circuits and observables (e.g., Pauli operators).
The sampler primitive generates samples / measurement outcomes (bitstrings) from circuits, typically with a specified number of shots. 
In Fig.~\ref{fig:qk-runtime}, we show a simple Qiskit program with the aforementioned
steps as well as a diagram illustrating the interaction between the Qiskit SDK and the Qiskit Runtime.
A code assistant should be fluent in understanding this workflow, and be able to make the right API calls given a problem and hardware or simulator requirements. For this to happen, it should be trained with high-quality code samples whose accuracy and execution can be verified on a quantum system. 

\subsection{Offline Interaction with a Quantum System}
\label{offline}
Given the interaction between the SDK and the runtime, we can design problems accompanied of unit tests
and ask an AI to generate code to answer the problems. The problem can be as simple as asking to connect to a particular
backend and return hardware specific parameters such as qubit relaxing time and decoherence time.
Or, we could also ask it for instance to design a variable quantum eigensolver (VQE) algorithm to solve a complex quantum chemistry problem.
The key for training the AI is to enable quantum verification using unit tests that guarantee that the job was
executed, and that the results returned are correct.

\begin{figure}[!b]
  \centering
\includegraphics[width=0.8\textwidth]{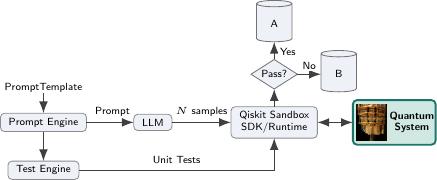}
  \caption{Generation pipeline for synthetic data.}
  \label{fig:synthetic-data}
\end{figure}

We developed a synthetic data pipeline that builds prompt/unit tests pairs, through templates that cover essential 
Qiskit features, for instance, estimators and samplers primitives, transpilation algorithms, or quantum error correction routines. 
Given the prompts, we use an LLM to generate code, and its answers are validated through 
execution of the code and the unit tests inside a sandbox environment with the latest Qiskit SDK and runtime installed.
Accepted and rejected code samples are saved in two buckets, defining databases that can be utilized for training the Qiskit Code Assistant.
Fig.~\ref{fig:synthetic-data} shows the full synthetic data generation process. Examples of generated data are presented in appendix~\ref{app:ex-synth-data}.
Note that not all the prompts require to interact with the runtime, and we also use quantum simulators such as the AerSimulator~\citep{javadiabhari2024quantumcomputingqiskit} which leverages real-backend noise models.
In practice, about 10\% of the problems we ran require access to the runtime.

After the generation process ends, the unit tests are discarded and we can use the prompts and the code for training an LLM.
The accepted data (bucket A) can directly be used for fine tuning. We can also build a dataset made of accepted and rejected (bucket B) responses for given prompts and post-train the LLM with Direct Preference Optimization (DPO)~\citep{rafailov2023direct}. 
   
\subsection{Online Interaction with a Quantum System}
\label{online}
\begin{figure}[!ht]
  \centering
\includegraphics[ width=0.8\textwidth]{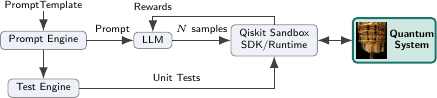}
  \caption{Training LLM through reinforcement learning from quantum verifiable rewards.}
  \label{fig:rlvr}
\end{figure}

We also propose a training approach where the LLM interacts directly with the quantum system 
through reinforcement learning, allowing it to improve its coding and reasoning skills using verifiable reward signals.
We use Group Relative Policy Optimization (GRPO) introduced in~\citep{shao2024deepseekmath} which is an online 
reinforcement learning algorithm that leverages grouped
 Monte Carlo rollouts to compute a normalized advantage estimate via a standardized reward.
In our setup, the reward is computed from the pass rate of the unit tests on the quantum system. It is formally defined as follows:
\begin{definitionbox}
A quantum verifiable reward is the percentage of unit tests that passed execution on a quantum hardware or quantum simulator.
\end{definitionbox}
Fig.~\ref{fig:rlvr} illustrates the training of a LLM through reinforcement learning from quantum verifiable rewards.
We use the same synthetic data pipeline as described previously to generate prompt/unit tests pairs.
During training, for each prompt, the LLM generates $N$ samples which are evaluated through the sandbox environment and the quantum system by executing the generated code together with the unit tests.
The sandbox returns the quantum verified reward which the GRPO algorithm uses to compute the advantage function and update the LLM's weights.

\section{Data and Model Training}
\label{sec:data-model}
\begin{figure}[!ht]
  \centering
  \includegraphics[width=0.9\textwidth]{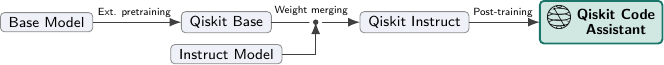}
  \caption{Training flowchart for the Qiskit Code Assistant.}
  \label{fig:qiskit-flow}
\end{figure}

We need a flexible and efficient training approach that enables models to generate correct code 
compatible with the latest version of the Qiskit SDK.
Indeed, whilst frontier models such as GPT-5~\citep{openai2025gpt5}, Claude-code~\citep{claude2025} or Gemini~\citep{gemini2025} all look reasonably good at writing
Qiskit programs, very often the Qiskit code they generate does not execute, or does not use the
preferred tools for solving a given quantum computing problem. Our training methodology involves 
three steps: (1) extended pretraining a base model using selected Qiskit code; 
(2) Model weight merging between the tuned base and an instruct model; and
(3) Model alignment on Qiskit problems. Fig.~\ref{fig:qiskit-flow} illustrates the training steps for building the Qiskit Code Assistant.

\subsection{Extended Pretraining and Weight Merging}

Our extended pretraining dataset\footnote{The data was crawled from GitHub and we use only the following licenses: Apache 2.0, MIT, the Unlicense, Mulan PSL Version 2, BSD-2, BSD-3, and Creative Commons Attribution 4.0.} consists of Qiskit Python code and Python notebooks, and we filter out deprecated and low-quality samples.
We also used Qiskit tutorials and Qiskit documentation to build question/answer pairs.
We used the Mixtral-8x7B-Instruct-v0.1~\citep{jiang2024mixtralexperts} model to generate the questions. For each document, the model was prompted to create general, unambiguous questions that could be understood without needing additional context. The same model then answered these questions using the original document. Finally, the model was given only the question-answer pairs (without access to the document) and asked to evaluate their quality.
The extended pretraining dataset also included Qiskit synthetic data built using the pipeline presented in Fig.~\ref{fig:synthetic-data}. Only the code passing
the execution (bucket A) is of interest for the extended pretraining corpus. Our full Qiskit extended pretraining corpus has around $82$~Million tokens. Table~\ref{tab:pt-data} in the appendix provides details on the token count and weight distribution for each subsets.

We extended the pretraining of \qwen~\citep{hui2024qwen2} on our Qiskit dataset, using one sample per 2048-token context window. We used LoRA with rank 64 and a scaling factor of 128. The model was trained for 2.5 epochs on 16 NVIDIA A100 80GB GPUs using a global batch size of 128.
The learning rate was linearly warmed up over 30 steps to reach $2\times10^{-4}$, after which it was annealed using a cosine schedule.
We then merged with \qweni using the Spherical Linear Interpolation (SLERP) method implemented in the open-source library mergekit~\citep{goddard-etal-2024-arcees} developed by Arcee AI.   
\subsection{Supervised Fine-Tuning and Distillation of Quantum Physics Knowledge}
Following the extended pretraining (EPT) and model merging, which produced a model specialized in Qiskit code generation, we aimed to expand its knowledge into the broader domain of quantum physics. The motivation for this step lies in the observation that effective solutions to Qiskit programming tasks often depend on a solid understanding of the underlying quantum principles. Such knowledge provides essential guidance for the correct formulation of Qiskit code.

To support this extension, we constructed a SFT dataset through a distillation process. We curated a batch of synthetic Qiskit problems, retaining only those that passed execution (bucket A in Fig.~\ref{fig:synthetic-data}). The DeepSeek V3 \citep{liu2024deepseek} model was then used as a teacher to provide explanatory descriptions of the quantum principles underlying each problem, while leaving aside the Qiskit programming details. The final dataset comprised triplets of: problem prompt, quantum explanation, and Qiskit solution. This dataset serves as a bridge between quantum concepts and code, allowing the model to internalize domain knowledge that improves its ability to interpret and solve Qiskit tasks.

We performed full fine-tuning of our EPT + weight-merged model on the generated synthetic dataset (combining quantum explanations and Qiskit code). The training was conducted for 3 epochs on 16 NVIDIA A100 80GB GPUs, with a batch size of 16. The learning rate was linearly warmed up over the first 90 steps to \(1 \times 10^{-5}\), after which it was decayed following a cosine schedule. In total, the supervised fine-tuning (SFT) dataset comprised 10{,}076 samples, where each instance was formatted with the quantum reasoning explanation delimited by <think> tokens, while the Qiskit code solution was placed within Python code blocks (\texttt{\`{}\`{}\`{}python}). During training, the loss was computed only on the completion tokens, with the prompt tokens excluded from gradient updates.

\subsection{Post-training Using Quantum Verification}
After SFT distillation, we further aligned the model using quantum verification. We tested two different post-training algorithms: DPO and GRPO. We also compared post-training on top of the SFT or directly on top of the EPT+instruct merged model.

For DPO, we built a synthetic dataset made of accepted and rejected responses using the process detailed in Fig.~\ref{fig:synthetic-data}.
For each prompt, we generate $N=16$ samples and randomly choose an accepted response.
If none is found, the prompt is discarded. 
We then used an embedding model to encode both the accepted and rejected responses, and selected the rejected sample that maximizes cosine similarity with the accepted one.
 The motivation for maximizing the similarity score is for the model
to learn subtle differences that can be hard to detect but nevertheless cause the code to fail when executing on the quantum system.
For the embedding model, we used codet5p-110m-embedding~\citep{wang2023codet5plus}. Our final dataset has around 4.9k rejected/accepted pairs.
We trained using LoRA with rank 64 and a scaling factor of 128. The model was trained for 3 epochs on 8 NVIDIA A100 80GB GPUs with a global batch size of 64. The learning rate was warmed up for 30 steps to $3\times10^{-5}$ and was then annealed following a cosine schedule. The $\beta$ temperature parameter controlling the sharpness of the preference weighting was set to $0.2$.

For GRPO, we built a synthetic prompt/unit test dataset with 4.5k samples. 
We trained all model weights using a learning rate of $1\times10^{-6}$ and a KL regularizer $\beta=0.01$. 
We used one reward based on quantum verification, and another formatting reward to force the model to reason and provide responses following a pre-defined format. The quantum verification reward had a weight of 1, and the format reward had a weight of 0.1. 
We used a group size $G=32$ and a global batch size of $512$. The maximum context length for the model was 1024, the maximum completion length was 2048 and the sampling temperature was  set  to $T=1.0$.
% We used the library TRL for the training using vLLM set in colocate mode.
% We set colocate mode, vLLM runs within the trainer process, sharing GPU memory with the training model which improves GPU utilization.
We trained for one epoch using 32 NVIDIA A100 80GB GPUs.

\section{Results}
\label{sec:results}
\paragraph{Evaluation benchmarks: QHE and QHE-hard.} We evaluated models using Qiskit-HumanEval (QHE)~\citep{qiskit_humaneval}, a dataset curated by Qiskit advocates whose goal is to measure the
ability of LLMs to solve Qiskit problems~\citep{vishwakarma2024qiskithumanevalevaluationbenchmark}. Across 151 problems, it covers a wild range of tasks including quantum circuit generation,
simulation and execution, algorithm implementations, circuit manipulation, circuit serialization, etc.
Similar to HumanEval~\citep{chen2021evaluatinglargelanguagemodels}, the LLM is to complete a python function, given its name, signature, and a docstring 
describing the problem to solve and the object(s) to return.
Each function is accompanied by executable unit tests, and the benchmark evaluates performance using \passatk~\citep{chen2021evaluatinglargelanguagemodels}.
Note that in this benchmark, most problems require to interact with the Qiskit runtime API, but only quantum simulators are used.
We also evaluate models on a modified version of QHE, Qiskit-HumanEval-hard (QHE-hard)~\cite{qiskit_humaneval_hard}.
In this variation, the prompt contains only the problem description, and the model is tasked with writing a function to solve it.
\emph{The main difference in QHE-hard with respect to  QHE is that the import statements
are not provided which makes the task more challenging, as the model needs to select which modules are the most appropriate to solve the problem.}

\begin{figure}[!ht]
  \centering
  \includegraphics[width=1.\textwidth]{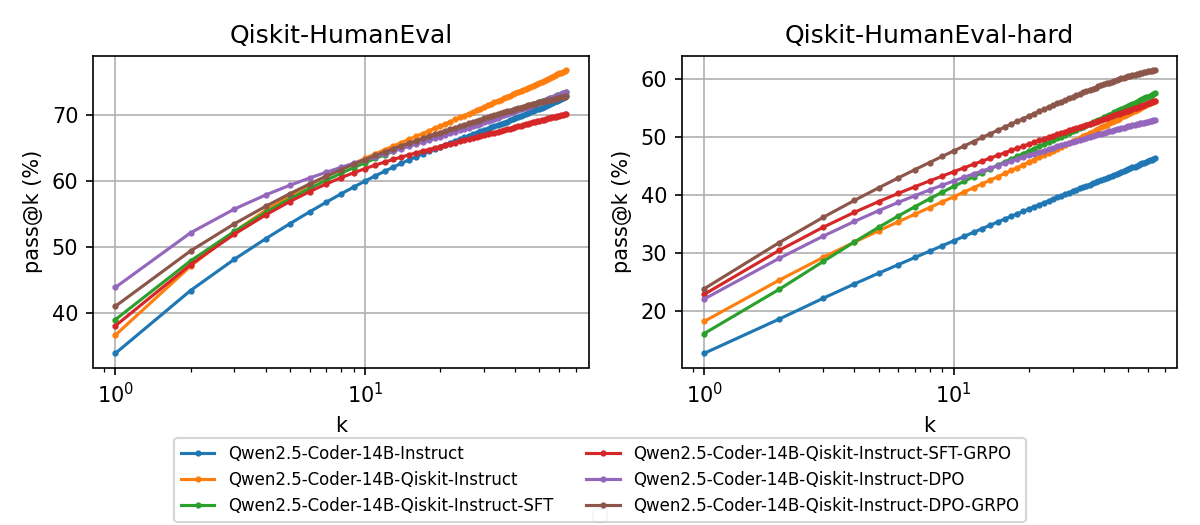}
  \caption{\passatk results on Qiskit-HumanEval and Qiskit-HumanEval-hard from $k=1$ to $k=64$. For clarity, we plot only the mean values here,
  see appendix~\ref{app:qhe-variance} for the variance.}
  \label{fig:qhe-sampling}
\end{figure}

\paragraph{Models and results.} We trained multiple models, ablating the SFT and post-training stages. Table~\ref{tab:stages} defines the models naming as well as the training stages for each.
Fig.~\ref{fig:qhe-sampling} shows results on QHE, and QHE-hard for the Qiskit models compared with \qweni. 
\begin{table}[t]
  \centering
  \caption{Training stages per Qiskit model.}
  \label{tab:stages}
  \begin{tabularx}{\linewidth}{p{7cm} *{5}{Y}}
    \toprule
    Model & EPT & Merge & SFT & DPO & GRPO \\
    \midrule
    \qwenqki & \tick & \tick & \blank & \blank & \blank \\
    \qwenqkis & \tick & \tick & \tick & \blank & \blank \\
    \qwenqkisg & \tick & \tick & \tick & \blank & \tick \\
    \qwenqkid & \tick & \tick & \blank & \tick & \blank \\
    \qwenqkidg & \tick & \tick & \blank & \tick & \tick \\
    \bottomrule
  \end{tabularx}
\end{table}
For each model, we sampled 64 sequences using a temperature $T=1$, and top-p equal to 1. On QHE, up to $k=20$, all the Qiskit models perform better than the original \qweni. 
At $k=1$, DPO and DPO-GRPO improve the pass rate of 8\%, and 5\% respectively, compared with \qwenqki.  
At $k=64$, the models reach the same accuracy around 73\%, with the exception of \qwenqki which is 4\% better and \qwenqkisg which is 3\% lower. In general, we observed that the SFT and SFT-GRPO models bring more background and analysis to the problems given, but tend to make more errors in coding.
On QHE-hard, all Qiskit models perform better than \qweni for all $k$, with a wide margin of 12-16\% for the best model. The Qiskit models do have a better understanding of Qiskit and solve more problems by using the right function and API calls. 
On QHE-hard, GRPO models outperform both instruct and DPO models, highlighting the benefits of reinforcement learning, which enables learning 
code executable on a quantum hardware without having access to the correct import statements.
Importing the correct modules is indeed learned by the model through self-play. 

Finally, we benchmarked the top open-source models on QHE and QHE-hard using greedy decoding. On QHE, the best-performing model is Qwen3-Coder-480B-A35B-Instruct, achieving 51.65\%. Our best Qiskit model reaches 50.33\%, slightly lower, but with $30\times$ fewer parameters. On QHE-hard, \qwenqkidg attains 28.48\%, the highest score among all evaluated models. We also confirmed that our Qiskit models maintain performance on HumanEval\citep{chen2021evaluatinglargelanguagemodels}, as shown by comparing \qweni with all Qiskit variants.

\begin{table}[h!]
\centering
\label{tab:benchmark-gd}
\caption{Model performance with greedy decoding on Qiskit-HumanEval (QHE), Qiskit-HumanEval-hard (QHE-hard) and HumanEval (HE).}
\begin{tabular}{lccc}
\toprule
\textbf{Model Name} & \textbf{QHE (\%)} & \textbf{QHE-hard (\%)} & \textbf{HE (\%)} \\
\midrule
gpt-oss-20b\textsuperscript{\dag}& 27.15 & 15.89 & 86.58 \\
Mistral-Small-3.2-24B-Instruct-2506 & 32.45 & 12.58 & 81.71 \\
Qwen3-14B\textsuperscript{\ddag} & 37.08 & 15.23 & 75.00 \\
\qweni\textsuperscript{\S} & 39.73 & 18.54 & 81.70 \\
Qwen3-Coder-30B-A3B-Instruct\textsuperscript{\ddag} & 45.69 & 19.21 & 90.85 \\
Qwen3-Coder-480B-A35B-Instruct\textsuperscript{\ddag} & \textbf{51.65} & 25.82 & \textbf{92.68} \\
Intern-S1\textsuperscript{\P} & 45.69 & 27.15 & \textbf{92.68} \\
\qwenqki & 45.03 & 22.51 & 89.02 \\
\qwenqkis  & 44.37 & 22.51 & 85.36 \\
\qwenqkid  & 50.33 & 26.49 & 89.02 \\
\qwenqkisg  & 46.35 & 26.49 & 86.58 \\
\qwenqkidg  & 45.69 & \textbf{28.48} & 88.41 \\
\bottomrule
\end{tabular}
\begin{flushleft}
\textsuperscript{\dag}~\citep{agarwal2025gpt}
\textsuperscript{\ddag}~\citep{qwen3technicalreport}
\textsuperscript{\S}~\cite{hui2024qwen2}
\textsuperscript{\P}~\citep{bai2025intern}
\end{flushleft}
\end{table}

\section{Related Work}
 Early work on neural networks for quantum computing~\citep{cruz2018deep} laid the groundwork for AI-driven approaches, while IBM’s Qiskit Code Assistant~\citep{dupuis2024qiskit} demonstrated that LLMs can be effectively specialized for quantum programming through domain-specific fine-tuning and evaluated with dedicated benchmarks~\citep{vishwakarma2024qiskithumanevalevaluationbenchmark}.
Building on these insights, researchers have explored diverse modeling strategies, such as treating quantum circuits as language tokens for generative modeling~\citep{nakaji2024generative}, integrating LLMs with quantum optimization~\citep{salloum2025q}, and employing multi-agent frameworks for iterative refinement~\citep{campbell2025enhancing}. Alternative generative approaches, including diffusion-based models, have also been proposed~\citep{beaudoin2025q}.
Complementary efforts have focused on scaling and evaluation, with large datasets of optimized circuits~\citep{jern2025fine}, specialized benchmarks like QHackBench~\citep{basit2025qhackbench}, and tools for taxonomy, refactoring, and defect prediction in quantum software~\citep{suarez2025taxonomy,suarez2025automatic,mao2025towards}. Practical innovations include lightweight local assistants~\citep{basit2025pennycoder}, comparative studies on gate design~\citep{closser2025pushing}, and reinforcement learning techniques to improve code generation reliability~\citep{kheiri2025qspark}.

\section{Conclusion}
We explored post-training techniques for LLMs to enhance quantum code generation and validated our approach using Qiskit code. Our results show that interacting with a quantum system is an effective strategy for producing high-quality Qiskit code and improving execution accuracy. To ensure correctness, we introduced 
\textit{quantum verification} as a mechanism to guarantee the quality and accuracy of generated code across diverse problem classes. By leveraging a quantum-verified data pipeline and testing both offline and online learning strategies, we achieved significant improvements on the Qiskit-HumanEval and Qiskit-HumanEval-hard benchmarks. Notably, a 14B-parameter model variant, fine-tuned on top of \qweni and post-trained with both DPO and GRPO, attained the highest score on Qiskit-HumanEval-hard, surpassing models with up to $30\times$ more parameters.

Whilst this work demonstrates that quantum verification can enhance LLM capabilities for solving Qiskit problems, it primarily focused on tasks of easy to intermediate complexity.
% To further enhance LLM reasoning and quantum intuition, more complex problems are needed. {\color{red}Add a comment about using fidelity, energy and other quantum aspects will be relevant also to solve more complex problems}
We believe that continued collaboration between the quantum and AI research communities is essential to develop challenging datasets and benchmarks that advance the capabilities of quantum code assistants.
In the era of AI-for-Quantum, we envision agentic workflows that can ultimately leverage quantum computers as tools for driving new scientific discoveries.

\bibliographystyle{abbrvnat}
\bibliography{references}

\newpage
\appendix
\UseRawInputEncoding

\section{Examples of Synthetic Data}
\label{app:ex-synth-data}
Below is an example of synthetic data built using the pipeline described in Fig.~\ref{fig:synthetic-data}. The prompt engine creates a synthetic prompt and the test engine creates unit tests. For fine-tuning, we use prompt/code pairs. For GRPO, we use the prompt, and the reward is calculated from the pass rate of the unit tests. For example if the model passed all the unit tests the reward is 1.  
\begin{tcolorbox}[colback=gray!5, rounded corners, boxrule=0.75pt,colframe=black]
\textbf{Prompt:}\\
Design a circuit named qc that has  2 quantum bits \&  2 classical bits. Then insert  controlled-rx gate between qubits 0 and 1, and set the gate parameters to theta=0.75. Next, generate a preset staged pass manager named PM using the ibm\_fez backend. Also, it should work with an optimization level 1. Now use it to transpile the circuit.
\vspace{0.5em}\\
\textbf{Code:}
\begin{tcolorbox}[colframe=black, sharp corners, boxrule=0.75pt]
\scalebox{0.8}{
\begin{lstlisting}[style=mypython,frame=none, numbers=none,]
from qiskit_ibm_runtime import QiskitRuntimeService
from qiskit.transpiler.preset_passmanagers import generate_preset_pass_manager
from qiskit import QuantumCircuit
from qiskit.circuit.library import CRXGate

# Create a circuit with 2 quantum bits and 2 classical bits
qc = QuantumCircuit(2, 2)

# Insert a controlled-RX gate between qubits 0 and 1 with theta=0.75
qc.append(CRXGate(0.75), [0, 1])

# Generate a preset staged pass manager using the ibm_fez backend and optimization level 1
service = QiskitRuntimeService()
backend = service.backend("ibm_fez")
PM = generate_preset_pass_manager(backend=backend, optimization_level=1)

# Transpile the circuit using the pass manager
transpiled_qc = PM.run(qc)
\end{lstlisting}
}
\end{tcolorbox}
\vspace{0.5em}
\textbf{Unit Tests:}
\begin{tcolorbox}[colframe=black, sharp corners, boxrule=0.75pt, ]
\scalebox{0.8}{
\begin{lstlisting}[style=mypython,frame=none, numbers=none,]
# synthetic unittests
import unittest
import pytest
class LLMGenTest(unittest.TestCase):
    def test_import_quantum_circuit__circuit__VarQuantumCircuit_0_num_qubits(self):
        assert qc.num_qubits == 2
    def test_import_quantum_circuit__circuit__VarQuantumCircuit_0_num_clbits(self):
        assert qc.num_clbits == 2
    def test_import_quantum_circuit__circuit__VarQuantumCircuit_0_gate_0_type(self):
        from qiskit.circuit import CircuitInstruction
        assert qc.data[0].operation.base_class == CircuitInstruction(CRXGate(0.75)).operation.base_class
    def test_import_quantum_circuit__circuit__VarQuantumCircuit_0_instruction_0_qubits_0(self):
        assert qc.data[0].qubits[0]._index == 0
    def test_import_quantum_circuit__circuit__VarQuantumCircuit_0_instruction_0_qubits_1(self):
        assert qc.data[0].qubits[1]._index == 1
    def test_import_quantum_circuit__circuit__VarQuantumCircuit_0_gate_0_params(self):
        assert qc.data[0].operation.params[:1] == [0.75]
    def test_import_gen_preset_pm__passmanager__VarGenPresetPM_0_pm_name(self):
        assert PM.__class__.__name__ == "StagedPassManager"
    def test_import_gen_preset_pm__passmanager__VarGenPresetPM_0_pm_optimization(self):
        assert {'InverseCancellation', 'Optimize1qGatesDecomposition'}.issubset({type(i).__name__ for i in PM.optimization.to_flow_controller().tasks[-1].tasks}) == True
if __name__ == '__main__':
    unittest.main()
\end{lstlisting}
}
\end{tcolorbox}
\end{tcolorbox}
\newpage

\section{Extended Pretraining Data Statistics}\label{app:pt-data}
Table~\ref{tab:pt-data} provides the token count and the weight attributed to each data subsets in the extended pretraining corpus. 
\begin{table}[htbp]
\centering
\caption{Qiskit Data used in extended pretraining.}
\small
\begin{tabularx}{\linewidth}{p{4cm} *{4}{Y}}
\toprule
\textbf{Dataset} &
\makecell{\textbf{Raw}\\\textbf{Tokens (M)}} &
\textbf{Epochs} &
\makecell{\textbf{Weight}\\\textbf{(\%)}} &
\makecell{\textbf{Eff. Tokens}\\\textbf{(M)}} \\
\midrule
\multicolumn{5}{l}{\textbf{Public GitHub Python}} \\
Preferred\textsuperscript{\dag} & 5.4 & 2.9 & 19.2 & \textbf{15.72} \\
Others & 57  & 1   & 69   & \textbf{57} \\
\addlinespace[0.5ex]
\multicolumn{5}{l}{\textbf{Public GitHub Notebooks}} \\
Preferred\textsuperscript{\dag} & 0.9 & 2.6 & 3    & \textbf{2.3} \\
Others & 4   & 1   & 4.8  & \textbf{4} \\
\addlinespace[0.5ex]
\multicolumn{5}{l}{\textbf{Qiskit Synthetic}} \\
Documentation            & 0.32 & 1 & 0.39 & \textbf{0.32} \\
Tutorial                 & 0.23 & 1 & 0.28 & \textbf{0.23} \\
Prompt/Code pairs & 2.6  & 1 & 3.16 & \textbf{2.6} \\
\addlinespace[0.5ex]
\bottomrule
\end{tabularx}
\vspace{2pt}
\footnotesize{\dag~Includes repositories from \href{https://github.com/qiskit}{Qiskit} and \href{https://github.com/qiskit-community}{Qiskit Community}.}
\label{tab:pt-data}
\end{table}

\section{\passatk Results on QHE and QHE-hard (with variance)}
\label{app:qhe-variance}
\begin{figure}[!ht]
  \centering
  \includegraphics[width=1.\textwidth]{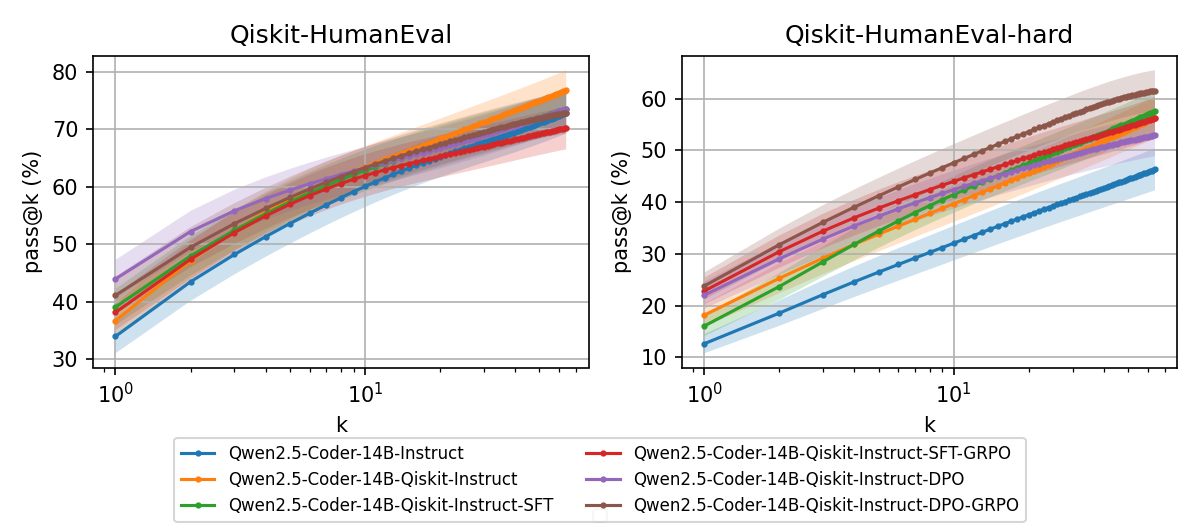}
  \caption{\passatk results on Qiskit-HumanEval and Qiskit-HumanEval-hard from $k=1$ to $k=64$. The shaded area represents $\mu\pm\sigma.$ }
\end{figure}

\section{Code Generation Examples}

\begin{tcolorbox}[colback=gray!5, rounded corners, boxrule=0.75pt,colframe=black, title={Qiskit DPO-GRPO vs. SFT-GRPO on Synthetic Prompt (Example 1)}]
\footnotesize
\tiny
\textbf{Prompt:}
\textit{\noindent Construct a random circuit called rand\_circ with  8 qbits and of depth  1. Also set measure to False. Remember to set its seed to 1. Once the circuit has been created generate a preset staged pass manager named PassManager using the FakeKyiv backend. Also it should be using optimization level 1. Finally use it to transpile the circuit.}
\begin{tcolorbox}[colframe=black, sharp corners, boxrule=0.75pt, colback=lightblue, width=\linewidth, left=2pt, right=2pt, top=2pt, bottom=0pt]
{\tiny
\textbf{Qwen2.5-Coder-14B-Qiskit-Instruct-DPO-GRPO}
\begin{lstlisting}[style=markdownwrap]
<think>
To construct the random circuit with the specified parameters and transpile it using the preset staged pass manager, I'll follow these steps:
1. Import the necessary modules from Qiskit.
2. Create the random circuit with 8 qubits and depth 1, setting the seed to 1 and disabling measurement.
3. Generate the preset staged pass manager for the FakeKyiv backend with optimization level 1.
4. Use the pass manager to transpile the circuit.
</think>
<answer>
```python
from qiskit.circuit.random import random_circuit
from qiskit.transpiler.preset_passmanagers import generate_preset_pass_manager
from qiskit_ibm_runtime.fake_provider import FakeKyiv
# Step 1: Create the random circuit
rand_circ = random_circuit(num_qubits=8, depth=1, seed=1, measure=False)
# Step 2: Generate the preset staged pass manager
backend = FakeKyiv()
PassManager = generate_preset_pass_manager(backend=backend, optimization_level=1)
# Step 3: Transpile the circuit using the pass manager
transpiled_circ = PassManager.run(rand_circ)
```
</answer>
\end{lstlisting}
}
\end{tcolorbox}
\begin{tcolorbox}[colframe=black, sharp corners, boxrule=0.75pt, colback=green!10!white, width=\linewidth, left=2pt, right=2pt, top=2pt, bottom=0pt]
{\tiny
\textbf{Qwen2.5-Coder-14B-Qiskit-Instruct-SFT-GRPO}
\begin{lstlisting}[style=markdownwrap]
<think>
To address this problem, we need to construct a quantum circuit with 8 qubits and a depth of 1, set the seed for reproducibility, and then transpile the circuit using a predefined staged pass manager with an optimization level of 1. Here’s the step-by-step reasoning process:
1. **Constructing the Random Circuit**:
   - We need to create a quantum circuit with 8 qubits. This means the circuit will have 8 quantum registers.
   - The depth of the circuit is set to 1, which implies that the circuit will consist of a single layer of quantum gates acting on the qubits. This typically means that each qubit will have at most one gate applied to it in this layer.
   - The `measure` parameter is set to `False`, indicating that no measurement operations will be included in the circuit.
   - The seed for the random number generator is set to 1 to ensure that the circuit is reproducible.
2. **Generating the Preset Staged Pass Manager**:
   - A staged pass manager is a sequence of optimization and transformation steps that are applied to a quantum circuit during the transpilation process. This is crucial for adapting the circuit to the specific constraints and capabilities of a quantum backend.
   - The `FakeKyiv` backend refers to a specific quantum hardware or simulator setup. Each backend has unique characteristics, such as qubit connectivity, gate set, and noise models, which influence how the circuit should be transpiled.
   - The optimization level of 1 indicates a moderate level of optimization. This typically involves basic circuit simplifications and optimizations without extensive gate decomposition or error mitigation.
3. **Transpiling the Circuit**:
   - Transpilation is the process of converting a quantum circuit into a form that is compatible with the target backend. This involves mapping the circuit’s gates to the backend’s native gate set, optimizing the circuit for efficiency, and ensuring that the circuit respects the backend’s physical constraints (e.g., qubit connectivity).
   - The staged pass manager, configured for the `FakeKyiv` backend with optimization level 1, will apply the necessary transformations to the circuit.
Given these steps, the final result will be a transpiled version of the random circuit that is optimized for the `FakeKyiv` backend, with the specified depth and qubit count.
</think>
<answer>
```python
from qiskit_ibm_runtime.fake_provider import FakeKyiv
from qiskit.circuit.random import random_circuit
from qiskit.transpiler.preset_passmanagers import generate_preset_pass_manager
# Create a random circuit with 8 qubits and depth 1
rand_circ = random_circuit(num_qubits=8, depth=1, measure=False, seed=1)
# Generate a preset PM using the FakeKyiv backend with opt. level 1, and transpile the circuit
PassManager = generate_preset_pass_manager(backend=FakeKyiv(), optimization_level=1)
transpiled_circ = PassManager.run(rand_circ)
```
</answer>
\end{lstlisting}
}
\end{tcolorbox}
\noindent\colorbox{yellow!30}{%
\parbox{0.95\linewidth}{\tiny
\textbf{Note:} The SFT model provides a more detailed explanation compared to DPO. 
Although both outputs are correct in this case, the SFT model provides richer 
quantum information about the prompt and its possible solution.}}
\end{tcolorbox}

\begin{tcolorbox}[colback=gray!5, rounded corners, boxrule=0.75pt,colframe=black, title={Qiskit DPO-GRPO vs. SFT-GRPO on Synthetic Prompt (Example 2)}]
\footnotesize
\tiny
\textbf{Prompt:}
\textit{\noindent Create an estimator named sim. Use FakeLimaV2 for the backend. Now create a quantum circuit qc with 5 qubits. Define the observables using SparsePauliOp in a variable named observable using 3 Pauli matrices with the following labels XXYYZ,XYZZI,XXZZZ, and set the coefficients to 1,-1,-1. Finally, run the estimator using a (circuit, observables, parameter\_values) tuple, the job should be called job.}
\vspace{0.5em}
\begin{tcolorbox}[colframe=black, sharp corners, boxrule=0.75pt, colback=lightblue, width=\linewidth, left=2pt, right=2pt, top=2pt, bottom=0pt]
{\tiny
\textbf{Qwen2.5-Coder-14B-Qiskit-Instruct-DPO-GRPO}
\begin{lstlisting}[style=markdownwrap]
<think>
To create an estimator named `sim` using FakeLimaV2 as the backend, I will first import the necessary modules from Qiskit. Then, I will define a quantum circuit with 5 qubits. After that, I will create observables using SparsePauliOp with the specified labels and coefficients. Finally, I will run the estimator with the defined circuit, observables, and parameter values to create a job named `job`.
</think>
<answer>
```python
from qiskit_ibm_runtime import EstimatorV2 as Estimator
from qiskit_ibm_runtime.fake_provider import FakeLimaV2
from qiskit.quantum_info import SparsePauliOp
from qiskit import QuantumCircuit
# Create an estimator with FakeLimaV2 backend
sim = Estimator(mode=FakeLimaV2())
# Create a quantum circuit with 5 qubits
qc = QuantumCircuit(5)
# Define the observables
observable = SparsePauliOp(["XXYYZ", "XYZZI", "XXZZZ"], coeffs=[1, -1, -1])
# Run the estimator
job = sim.run([(qc, observable)])
```
</answer>
\end{lstlisting}
}
\end{tcolorbox}
\begin{tcolorbox}[colframe=black, sharp corners, boxrule=0.75pt, colback=green!10!white, width=\linewidth, left=2pt, right=2pt, top=2pt, bottom=0pt]
{\tiny
\textbf{Qwen2.5-Coder-14B-Qiskit-Instruct-SFT-GRPO}
\begin{lstlisting}[style=markdownwrap]
<think>
To create an estimator named `sim` using `FakeLimaV2` as the backend, we need to follow a series of steps that involve constructing a quantum circuit, defining observables, and running an estimation job. Let's break this down step by step.
1. **Backend Initialization**: The backend `FakeLimaV2` is a simulated quantum device that mimics the behavior of a real quantum computer. It is used here to perform the quantum computations in a controlled environment.
2. **Quantum Circuit Creation**: We need to create a quantum circuit `qc` with 5 qubits. This circuit will be the quantum system on which we will apply operations and measure observables.
3. **Observable Definition**: The observables are defined using the `SparsePauliOp` class, which represents operators in the Pauli basis. We are given three Pauli strings: "XXYYZ", "XYZZI", and "XXZZZ". These strings correspond to the Pauli matrices (I, X, Y, Z) acting on the respective qubits. The coefficients for these observables are set to 1, -1, and -1, respectively. These coefficients determine the weights of the observables in the final expectation value calculation.
4. **Estimator Execution**: The estimator will compute the expectation values of the defined observables with respect to the quantum state produced by the circuit. This involves running the circuit and measuring the observables. The results will be stored in a job, which we will call `job`.
5. **Reasoning Process**: 
   - The `SparsePauliOp` is a way to represent Hamiltonians or observables in terms of Pauli operators, which are fundamental in quantum mechanics. The Pauli strings "XXYYZ", "XYZZI", and "XXZZZ" are specific combinations of Pauli-X and Pauli-Z operators acting on different qubits.
   - The coefficients (1, -1, -1) indicate that we are considering a weighted sum of these observables. The expectation value calculated by the estimator will be a linear combination of the expectation values of these observables.
   - The backend `FakeLimaV2` simulates the quantum system, allowing us to perform these calculations in a virtual environment.
6. **Final Output**: The `job` will contain the results of the estimation, including the expectation values of the observables. These results are crucial for understanding the behavior of the quantum system described by the circuit.
</think>
<answer>
```python
from qiskit_ibm_runtime.fake_provider import FakeLimaV2
from qiskit_ibm_runtime import EstimatorV2 as Estimator
from qiskit.quantum_info import SparsePauliOp
from qiskit import QuantumCircuit
# Initialize the backend
backend = FakeLimaV2()
# Create an estimator using the backend
sim = Estimator(mode=backend)
# Create a quantum circuit with 5 qubits
qc = QuantumCircuit(5)
# Define the observables
observable = SparsePauliOp(["XXYYZ", "XYZZI", "XXZZZ"], coeffs=[1, -1, -1])
# Run the estimator
job = sim.run([(qc, observable)])
```
</answer>
\end{lstlisting}
}
\end{tcolorbox}
\end{tcolorbox}

\begin{tcolorbox}[colback=gray!5, rounded corners, boxrule=0.75pt,colframe=black, title={Wrong Code Generation Case for a Synthetic Prompt (Qiskit DPO-GRPO, also fails with Qiskit GRPO-SFT)}]
\footnotesize
\tiny
\textbf{Prompt:}
\begin{itshape}
\noindent Define an estimator called est using the ibm\_fez backend. Now, I would like to enable ZNE for it. Configure the resilience options with the following settings: enable zne mitigation, set noise factors to [1.0, 2.0, 4.0], choose extrapolation method polynomial\_degree\_2.
\end{itshape}
\vspace{0.5em}
\begin{tcolorbox}[colframe=black, sharp corners, boxrule=0.75pt, colback=lightred]
{\tiny
\textbf{Qwen2.5-Coder-14B-Qiskit-Instruct-DPO-GRPO}
\begin{lstlisting}[style=markdownwrap]
<think>
To define an estimator and enable Zero Noise Extrapolation (ZNE) with the specified settings on the `ibm_fez` backend, I need to follow these steps:

1. Import the necessary modules from Qiskit.
2. Initialize the estimator with the `ibm_fez` backend.
3. Configure the resilience options of the estimator to enable ZNE and set the specified parameters.

Let's go through each step in detail.
</think>
<answer>
```python
from qiskit_ibm_runtime import EstimatorV2 as Estimator
from qiskit_ibm_runtime import QiskitRuntimeService

# Step 1: Define the estimator using ibm_fez backend
service = QiskitRuntimeService()
ibm_fez = service.backend("ibm_fez")
est = Estimator(mode=ibm_fez)

# Step 2: Enable ZNE and set the resilience options
est.options.resilience_level = 1  # Enable ZNE mitigation
est.options.resilience.zne.noise_factors = [1.0, 2.0, 4.0]
est.options.resilience.zne.extrapolator = "polynomial_degree_2"
```

This code snippet initializes an estimator named `est` using the `ibm_sherbrooke` backend and configures it with Zero Noise Extrapolation enabled, setting the noise factors to `[1.0, 2.0, 4.0]` and the extrapolation method to `polynomial_degree_2`.
</answer>
\end{lstlisting}
}
\end{tcolorbox}
\noindent\colorbox{yellow!30}{%
\parbox{0.95\linewidth}{\tiny
\textbf{Note:} During the GRPO training, this code achieved a reward of 0.8. Although it is mostly correct, the method used to enable Zero-Noise Extrapolation (ZNE) was incorrect. Specifically, setting \texttt{resilience\_level = 1} does not automatically activate ZNE. Instead, it must be explicitly enabled by setting \texttt{est.options.resilience.zne\_mitigation = True}.
}}
\end{tcolorbox}

\begin{tcolorbox}[colback=gray!5, rounded corners, boxrule=0.75pt,colframe=black, title={Qiskit DPO-GRPO vs. SFT-GRPO on Non-Synthetic Prompt (Example 1)}]
\footnotesize
\tiny
\textbf{Prompt:}
\textit{\noindent Show me how to define a ghz state in qiskit. Then I want you to convert it into a qasm3 string}
\vspace{0.5em}
\begin{tcolorbox}[colframe=black, sharp corners, boxrule=0.75pt, colback=lightblue, width=\linewidth, left=2pt, right=2pt, top=2pt, bottom=0pt]
{\tiny
\textbf{Qwen2.5-Coder-14B-Qiskit-Instruct-DPO-GRPO}
\begin{lstlisting}[style=markdownwrap]
<think>
To define a GHZ (Greenberger-Horne-Zeilinger) state in Qiskit and convert it into a QASM3 string, we will follow these steps:
1. Import the necessary modules from Qiskit.
2. Create a quantum circuit to represent the GHZ state.
3. Apply the necessary quantum gates to prepare the GHZ state.
4. Convert the quantum circuit to a QASM3 string.
Let's go through these steps in detail.
</think>
<answer>
Here's how you can define a GHZ state in Qiskit and convert it into a QASM3 string:
```python
from qiskit import QuantumCircuit
from qiskit.circuit.library import HGate, CXGate
from qiskit.qasm3 import dumps
# Create a quantum circuit with 3 qubits and 3 classical bits for measurement
qc = QuantumCircuit(3, 3)
# Apply H gate to the first qubit to create a superposition
qc.append(HGate(), [0])
# Apply CX gates to entangle the qubits
qc.append(CXGate(), [0, 1])
qc.append(CXGate(), [0, 2])
# Measure all qubits
qc.measure(range(3), range(3))
# Convert the quantum circuit to a QASM3 string
qasm3_string = dumps(qc)
# Print the QASM3 string
print(qasm3_string)
```
This code will create a GHZ state on three qubits and then convert the resulting quantum circuit into a QASM3 string, which you can use for further processing or visualization.
</answer>
\end{lstlisting}
}
\end{tcolorbox}
\begin{tcolorbox}[colframe=black, sharp corners, boxrule=0.75pt, colback=green!10!white, width=\linewidth, left=2pt, right=2pt, top=2pt, bottom=0pt]
{\tiny
\textbf{Qwen2.5-Coder-14B-Qiskit-Instruct-SFT-GRPO}
\begin{lstlisting}[style=markdownwrap]
<think>
To define a GHZ (Greenberger-Horne-Zeilinger) state in Qiskit and convert it into a QASM3 string, we need to follow these steps:
1. **Define the GHZ State**: A GHZ state is a quantum state that is entangled across multiple qubits. For \( n \) qubits, the GHZ state is given by:
   \[
   |\text{GHZ}\rangle = \frac{1}{\sqrt{2}}(|0\rangle^{\otimes n} + |1\rangle^{\otimes n})
   \]
   This state is created by applying a Hadamard gate to the first qubit and then applying a series of CNOT gates between the first qubit and each of the subsequent qubits.
2. **Implement the GHZ State in Qiskit**: We will use Qiskit's quantum circuit framework to create a circuit that prepares the GHZ state. The circuit will consist of a Hadamard gate on the first qubit and CNOT gates between the first qubit and each of the other qubits.
3. **Convert the Circuit to QASM3**: Once the circuit is defined, we can use Qiskit's functionality to convert the circuit into a QASM3 string. QASM3 is a quantum assembly language that describes quantum circuits in a standardized format.
Let's go through the implementation step-by-step.
</think>
<answer>
```python
from qiskit.circuit.library import HGate, CXGate
from qiskit import QuantumCircuit
from qiskit.qasm3 import dumps

# Define the number of qubits
n_qubits = 3

# Create a quantum circuit with n_qubits
qc = QuantumCircuit(n_qubits)

# Apply Hadamard gate to the first qubit
qc.append(HGate(), [0])

# Apply CNOT gates between the first qubit and each of the other qubits
for i in range(1, n_qubits):
    qc.append(CXGate(), [0, i])

# Convert the circuit to QASM3 string
qasm3_string = dumps(qc)

print(qasm3_string)
```

This code snippet defines a GHZ state for 3 qubits and converts the corresponding quantum circuit into a QASM3 string. The output will be a string representation of the quantum circuit in QASM3 format.
</answer>
\end{lstlisting}
}
\end{tcolorbox}
\end{tcolorbox}

\scalebox{0.96}{%
\begin{tcolorbox}[colback=gray!5, rounded corners, boxrule=0.75pt,colframe=black, title={Qiskit DPO-GRPO vs. SFT-GRPO on Non-Synthetic Prompt (Example 2)}]
\footnotesize
\tiny
\textbf{Prompt:}
\textit{\noindent Give me the ibm backends with more than 5 qubits that are operational and picks the least busy one. Then create a Bell circuit and execute the circuit on the selected backend with a specified number of shots using the sampler primitive.}
\begin{tcolorbox}[colframe=black, sharp corners, boxrule=0.75pt, colback=lightblue, width=\linewidth, left=2pt, right=2pt, top=2pt, bottom=0pt]
{\tiny
\textbf{Qwen2.5-Coder-14B-Qiskit-Instruct-DPO-GRPO}
\begin{lstlisting}[style=markdownwrap, basicstyle=\ttfamily\tiny]
<think>
To accomplish this task, I'll follow these steps:
1. Import necessary modules from Qiskit.
2. Use the IBM Quantum provider to get a list of backends.
3. Filter the backends to select only those with more than 5 qubits that are operational.
4. Identify the least busy backend from the filtered list.
5. Create a Bell circuit.
6. Transpile the Bell circuit.
7. Use the sampler primitive to execute the circuit on the selected backend with a specified number of shots.
</think>
<answer>
```python
from qiskit_ibm_runtime import QiskitRuntimeService, SamplerV2 as Sampler
from qiskit import QuantumCircuit
from qiskit.transpiler.preset_passmanagers import generate_preset_pass_manager
from qiskit.circuit.library import HGate, CXGate
# Initialize QiskitRuntimeService
service = QiskitRuntimeService()
# Filter backends with more than 5 qubits that are operational
filtered_backends = [backend for backend in service.backends() if backend.configuration().n_qubits > 5 and backend.status().operational]
# Sort the filtered backends by least busy
filtered_backends.sort(key=lambda backend: backend.status().pending_jobs)
# Select the least busy backend
least_busy_backend = filtered_backends[0]
# Create a Bell circuit
bell = QuantumCircuit(2, 2)
bell.append(HGate(), [0])
bell.append(CXGate(), [0, 1])
bell.measure(range(2), range(2))
# Transpile the Bell circuit
transpiled_bell = generate_preset_pass_manager(backend=least_busy_backend).run(bell)
# Initialize the Sampler primitive
sampler = Sampler(least_busy_backend)
# Execute the circuit on the selected backend with 1024 shots
job = sampler.run([transpiled_bell])
```
</answer>
\end{lstlisting}
}
\end{tcolorbox}
\begin{tcolorbox}[colframe=black, sharp corners, boxrule=0.75pt, colback=green!10!white, width=\linewidth, left=2pt, right=2pt, top=2pt, bottom=0pt]
{\tiny
\textbf{Qwen2.5-Coder-14B-Qiskit-Instruct-SFT-GRPO}
\begin{lstlisting}[style=markdownwrap, basicstyle=\ttfamily\tiny]
<think>
To address this problem, we need to break it down into several steps:
1. **Filter IBM Quantum Backends**: We need to filter the available IBM Quantum backends to select only those with more than 5 qubits that are operational. This involves querying the IBM Quantum service for backend information and filtering based on the number of qubits and operational status.
2. **Select the Least Busy Backend**: From the filtered list, we need to select the backend that is currently the least busy. This involves checking the current usage or queued jobs of each backend to determine which one is least occupied.
3. **Create a Bell Circuit**: A Bell circuit is a simple quantum circuit that creates entanglement between two qubits. This involves preparing a Bell state, which is a superposition of the basis states \(|\psi\rangle = \frac{1}{\sqrt{2}}(|00\rangle + |11\rangle)\).
4. **Transpile the Circuit**: Transpilation is the process of converting a quantum circuit into a form that is compatible with the specific backend. This includes mapping the logical qubits to physical qubits and optimizing the circuit for the backend's constraints.
5. **Execute the Circuit Using the Sampler Primitive**: The sampler primitive is a tool for sampling from the outcome distribution of a quantum circuit. This involves running the circuit on the selected backend and obtaining the measurement results.
</think>
<answer>
```python
from qiskit.circuit.library import HGate, CXGate
from qiskit import QuantumCircuit
from qiskit.transpiler.preset_passmanagers import generate_preset_pass_manager
from qiskit_ibm_runtime import QiskitRuntimeService, SamplerV2 as Sampler
# Set up the IBM Quantum service
service = QiskitRuntimeService()
# Filter backends
backends = service.backends(simulator=False, operational=True)
filtered_backends = [backend for backend in backends if backend.configuration().n_qubits > 5]
# Select the least busy backend
least_busy_backend = min(filtered_backends, key=lambda b: b.status().pending_jobs)
# Create a Bell circuit
bell_circuit = QuantumCircuit(2)
bell_circuit.append(HGate(), [0])
bell_circuit.append(CXGate(), [0, 1])
bell_circuit.measure_all()
# Transpile the circuit
transpiled_circuit = generate_preset_pass_manager(backend=least_busy_backend).run(bell_circuit)
# Execute the circuit using the sampler primitive and get the result
sampler = Sampler(mode=least_busy_backend)
job = sampler.run([transpiled_circuit], shots=1024)
result = job.result()
```
</answer>
\end{lstlisting}
}
\end{tcolorbox}
\end{tcolorbox}
}

\section{Assets}
\subsection{Hardware For Training and Evaluating Models}
Our experiments were run on Vela, an AI-optimized supercomputing infrastructure integrated into the Cloud
and leveraging the RedHat OpenShift platform. 
Each node of the cluster has eight 80GB NVIDIA A100 GPUs, connected to each other by NVLink and NVSwitch. All the nodes have 3rd Generation Intel Xeon Scalable processors (Ice Lake), 1.5TB of DRAM, and four 3.2TB NVMe drives. More detail on the platform can be found in~\citep{gershon2025infrastructurepoweringibmsgen}.

\subsection{Libraries}
We used open-source libraries \href{https://pytorch.org/}{\texttt{pytorch}} \citep{Paszke2019} (license: BSD), \href{https://github.com/huggingface/transformers}{\texttt{HuggingFace Transformers}} \citep{Wolf2020} (Apache 2.0 license), and \href{https://github.com/huggingface/trl}{\texttt{HuggingFace TRL}} \citep{Vonwerra2022} (Apache 2.0 license).

\subsection{Code re-use}
Our GRPO training code is based on the public GitHub repository \url{https://github.com/huggingface/open-r1} \citep{Openr12025}.

\end{document}